\documentclass[a4paper, 12pt]{article}
\usepackage{amssymb}
\usepackage{amsmath,amsfonts,epsfig,subfigure}

\begin{document}

\title{Rayleigh waves and surface stability \\
for Bell materials in compression;\\
    comparison with rubber}

\author{Michel Destrade}

\date{2003}

\maketitle

\bigskip

\begin{abstract}
\noindent
The stability of a Bell-constrained half-space in compression is 
studied.
To this end, the propagation of Rayleigh waves on the 
surface of the material when it is maintained in a static state of 
triaxial prestrain is considered. 
The prestrain is such that the free surface of the half-space is
a principal plane of deformation.
The exact secular equation is established for surface waves traveling 
in a principal direction of strain with attenuation along the 
principal direction normal to the free plane. 
As the half-space is put under increasing compressive loads, the 
speed of the wave eventually tends to zero and the bifurcation 
criterion, or stability equation, is reached.

Then the analysis is specialized to specific forms of strain energy
functions and prestrain, and comparisons are made with results 
previously obtained in the case of incompressible neo-Hookean or 
Mooney-Rivlin materials. 
It is found that these rubber-like incompressible materials may
be compressed more than ``Bell empirical model'' materials, but not as 
much as ``Bell simple hyperelastic'' materials, before the critical 
stretches, solutions to the bifurcation criterion, are reached.
In passing, some classes of incompressible materials which possess a
relative-universal bifurcation criterion are presented.

\end{abstract}

\newpage

\section{Introduction}

The works of Maurice Anthony Biot (1905-1985) cover a wide range of 
topics in mechanics and applied mathematics.
Although much attention has been devoted to his contributions to the 
``acoustics, elasticity, and thermodynamics of porous media'' 
\cite{Biot91}, his results in finite and incremental elasticity 
\cite{Biot65} were also far-reaching and are still relevant to many 
contemporary problems.
For instance, he wrote a series of articles (summarized in his 
textbook \cite{Biot65}) on the surface and interfacial instability 
of elastic media under compression and his results found applications
in rubber elasticity, viscoelasticity, folding of 
inhomogeneous/multilayered media, geological structures, etc.
The idea underlying his resolution of these problems is the
following:
consider a media at rest under a finite compression;
superpose an incremental inhomogeneous static deformation whose 
amplitude vanishes away from the interface;
show that the initial compression leads to an interface deflection 
which is infinite;
conclude that this condition corresponds to interface bulking or 
instability.
Biot also noted that the dynamical counterparts to surface and 
interface stability analyses were Rayleigh and Stoneley wave 
propagation, respectively.

This paper studies the propagation of Rayleigh waves on the surface
of a compressed, internally constrained, hyperelastic half-space.
The corresponding ``bulking'' or ``bifurcation'' criterion is
derived by determining under which compressive loads the wave
speed tends to zero in the secular equation.
Biot often considered materials subject to incompressibility, 
an internal constraint which in nonlinear elasticity imposes that
$\text{det }\mathbf{V}=1$ at all times, where $\mathbf{V}$ is the
left stretch tensor.
Here, the materials considered are subject to the constraint of Bell
\cite{BeHa92a}, $\text{tr }\mathbf{V}=3$.
Both constraints are equivalent in infinitesimal linear elasticity
(they reduce to: $\text{tr }\mathbf{E}=0$, where $\mathbf{E}$ is the
infinitesimal strain tensor) but lead to quite different results at 
finite strains.
In particular, the secular equation for Rayleigh surface waves cannot 
be deduced for Bell materials from the incompressible case.
This equation was obtained in Ref.\cite{Dest02b} as a cubic for the 
squared wave speed.
However, this cubic corresponds to the rationalization of the exact 
secular equation and has spurious roots \cite{Rome01};
accordingly, the corresponding relevant bifurcation criterion would 
have to be carefully selected.
Here the \textit{exact} secular equation is found, as well as the 
\textit{exact} bifurcation criterion.

For incompressible materials, the Mooney-Rivlin form of the strain 
energy function $\Sigma_{\text{MR}}$ brings satisfactory correlation 
between theory and experiments for rubber-like materials;
this function $\Sigma_{\text{MR}}$ is linear with respect to $I_1$ and 
$I_2$, the first and second invariants of the left Cauchy-Green strain 
tensor $\mathbf{B} = \mathbf{V}^2$.
For Bell constrained materials, the strain energy density for  
``simple hyperelastic Bell materials'' \cite{BeHa92a} is linear with 
respect to $i_2$ and $i_3$, the second and third invariants of 
$\mathbf{V}$.
Regarding experimental results, the strain energy
function for ``Bell empirical model'' \cite{Bell85}, 
$\Sigma_{\text{BEM}}=(2/3)\beta_0 [2(3-i_2)]^{\textstyle{\frac{3}{4}}}$
(where $\beta_0$ is a material constant) is reported as consistent
with many trials on annealed metals such as Aluminum, Copper, or Zinc.
After the equations governing the problem have been written and 
solved in Section 2 for a general form of the strain energy function
for a Bell-constrained half-space,
the analysis is specialized in Section 3 to the two specific forms of 
strain energy functions presented above, and the results are compared
to those obtained by Biot for rubber-like materials.
It turns out that the maximal compressive load that can be applied 
to a half-space before the bifurcation criterion is reached 
is larger (smaller) for simple hyperelastic Bell materials 
(Bell empirical model) than for Mooney-Rivlin incompressible 
materials.
Also, the bifurcation criterion is the same for every material within
each class, and an infinity of strain energy densities for which 
incompressible half-spaces admit such  ``universal'' bifurcation
criteria is presented in 
\S\ref{relative-universal-bifurcation-criteria}.
Finally in Section 4, the pertinence of the notion of (in)stability 
for finitely deformed hyperelastic materials is briefly reviewed and 
the general interest of the Bell constraint is discussed, as opposed 
to the constraint of incompressibility.

\section{Resolution of the problem in the general case}
%
\subsection{Finite pure homogeneous triaxial pre-stretch}
%
Let ($O, x_1,x_2,x_3$) $\equiv$
($O,\mathbf{i}, \mathbf{j}, \mathbf{k}$) be a Cartesian rectangular
coordinate system.
Let the half-space $x_2 \ge 0$ be occupied by a hyperelastic
Bell-constrained material, with strain energy density $\Sigma$.
This material is subject to the internal constraint that for any
deformation  \cite{BeHa92a,BeHa92b},
\begin{equation} \label{Bell}
i_1 \equiv \text{tr } \mathbf{V} =3,
\end{equation}
at all times, where $\mathbf{V}$ is the left stretch tensor.
Hence, for isotropic Bell materials, $\Sigma$ depends only upon
$i_2$ and $i_3$, the respective second and third invariants of
$\mathbf{V}$.
So, $\Sigma = \Sigma(i_2,i_3)$, where
\begin{equation}
i_2 = [ (\text{tr } \mathbf{V})^2 - \text{tr } (\mathbf{V}^2)]/2, \:
i_3 = \text{det } \mathbf{V},
\end{equation}
and the constitutive equation giving the Cauchy stress tensor
$\mathbf{T}$ is \cite{BeHa92a}
\begin{equation} \label{constitutive}
 \mathbf{T}= p \mathbf{V} + \omega_0 \mathbf{1}
  +  \omega_2 \mathbf{V}^2,
\end{equation}
where $p$ is an arbitrary scalar, to be found from the equations of
motion and the boundary conditions; and the material response
functions $ \omega_0$ and $ \omega_2$ are defined by
\begin{equation} \label{omega}
\omega_0 = \partial \Sigma/\partial i_3,
\quad
\omega_2 = -i^{-1}_3 \partial \Sigma/\partial i_2,
\end{equation}
and should verify the Beatty--Hayes  $A$-inequalities \cite{BeHa92a}
\begin{equation} \label{A-inequalities}
\omega_0 (i_2, i_3) \le 0, \quad \omega_2 (i_2, i_3) > 0.
\end{equation}

In the case where the material is maintained in a state of finite
pure homogeneous static deformation, with principal stretch ratios
$\lambda_1, \lambda_2, \lambda_3$, along the $x_1,x_2,x_3$, axes,
the Cauchy stress tensor is the constant tensor $\mathbf{T}_o$ given
by
\begin{equation}  \label{T}
\mathbf{T_{o}} = (p_o \lambda_1 + \omega_0
  + \lambda_1^2 \omega_2) \mathbf{i} \otimes \mathbf{i} +
(p_o \lambda_2 + \omega_0
  + \lambda_2^2 \omega_2) \mathbf{j} \otimes \mathbf{j} +
(p_o \lambda_3 + \omega_0
  + \lambda_3^2 \omega_2) \mathbf{k} \otimes \mathbf{k}.
\end{equation}
Here $ \omega_0$ and $ \omega_2$ are evaluated at $i_2$, $i_3$
given by
\begin{equation} \label{invariants-static}
i_2 = \lambda_1 \lambda_2+ \lambda_2 \lambda_3+ \lambda_3 \lambda_1,
\quad
i_3 = \lambda_1 \lambda_2 \lambda_3.
\end{equation}
Of course, 
\begin{equation} \label{BellLambdas}
\lambda_1+ \lambda_2+ \lambda_3 = 3,
\end{equation} 
in order to satisfy \eqref{Bell}.
It is assumed that the boundary $x_2=0$ is free of 
tractions so that $T_{o22}=0$; 
and that the compressive loads $P_1$ and $P_3$ are applied at 
$x_1=\infty$ and $x_3=\infty$ to maintain the deformation, so that 
$P_1 = - T_{o11}$ and $P_3 = -T_{o33}$.
Hence,
\begin{equation} \label{po}
p_o = - ( \omega_0   + \lambda_2^2 \omega_2)/ \lambda_2,
\quad
P_\Gamma =  (\lambda_2 -  \lambda_\Gamma)
    (-\omega_0 + \lambda_\Gamma \lambda_2 \omega_2)/ \lambda_2,
\quad (\Gamma =1,3).
\end{equation}

%
\subsection{Incremental equations for surface waves}
%
Beatty and Hayes \cite{BeHa95b} wrote the general equations for 
small-amplitude motions in a Bell-constrained 
material maintained in a static state of finite pure homogeneous 
deformation (as described in the previous subsection).
These equations were then specialized by this author \cite{Dest02b} 
to surface (Rayleigh) waves.
The infinitesimal superposed wave is of the form
$\Re \{ \mathbf{U}(kx_2) e^{i k (x_1 - vt)} \}$, 
where $\mathbf{U}$ is an unknown decaying function.
Hence the wave propagates in the direction of the $x_1$-axis with
speed $v$ and wave number $k$ and is attenuated in the direction of
the $x_2$-axis.
The incremental tractions acting upon the planes 
$x_2=$ const. are $\sigma^*_{21}$ and $\sigma^*_{22}$, 
and the introduction of the scalars functions 
$t_1(kx_2)$ and $t_2(kx_2)$, defined by
\begin{equation} \label{sigma}
\sigma^*_{21}(x_1,x_2,t) = kt_1(k x_2)e^{ik(x_1 - vt)}, \quad
\sigma^*_{22}(x_1,x_2,t) = kt_2(k x_2)e^{ik(x_1 - vt)},
\end{equation}
allows for a compact and simple form of the equations of motion and 
of the boundary conditions.
Explicitly, the equations of motion are \cite{Dest02b}
\begin{align} \label{1stOrder}
& t_1' + i \lambda_1 \lambda_2^{-1} t_2
    - (\lambda_1 \lambda_2^{-1}C - \rho v^2) U_1 = 0,  \nonumber 
\\
& t_2' + i  t_1 
        -[b_3(\lambda_1^2 -  \lambda_2^2) - \rho v^2 ] U_2= 0, 
\nonumber 
\\
& U_2' + i \lambda_1 \lambda_2^{-1} U_1 = 0,  
\\
& b_3 \lambda_2^2 U_1' + ib_3 \lambda_2^2 U_2 -  t_1 = 0. \nonumber
\end{align}
Here the prime denotes differentiation with respect to $k x_2$, 
$\rho$ is the mass density of the material, and 
\begin{align} \label{coefficients}
& b_3 = \frac{-\omega_0 + \lambda_1 \lambda_2 \omega_2} 
        {\lambda_2(\lambda_1 + \lambda_2)} >0,
\nonumber 
\\
& C_{\alpha \beta}  = 
   2 \lambda^2_{\alpha} \delta_{\alpha \beta} \omega_2
    - \lambda_{\beta}^2 (\omega_{02} + \lambda_{\alpha}^2 \omega_{22}) 
    + \lambda_1 \lambda_2  \lambda_3 (\omega_{03} + 
                          \lambda_{\alpha}^2 \omega_{23}),
\\
& C = \lambda_1^{-1} \lambda_2 C_{11}
    + \lambda_1 \lambda_2^{-1} C_{22} -C_{12} - C_{21} 
       -2\omega_0 - (\lambda_1^2+\lambda_2^2)\omega_2,
\nonumber 
\end{align}
where the derivatives $\omega_{0\Gamma}$, $\omega_{2\Gamma}$
($\Gamma = 2,3$) of the material response functions $\omega_0$,
$\omega_2$ are taken with respect to $i_\Gamma$ and evaluated at $i_2$,
$i_3$ given by \eqref{invariants-static}.
Note that the quantity $b_3$ defined above is positive according to 
the $A$-inequalities \eqref{A-inequalities}.
Finally, the boundary conditions are simply
\begin{equation}    \label{BC1}
t_1(0) = t_2(0) = 0.
\end{equation}

\subsection{Exact secular equation and exact bifurcation criterion}

The incremental Bell constraint Eq.\eqref{1stOrder}$_3$
suggests the introduction of a function $\varphi$ defined by
\begin{equation} \label{phiHat}
U_1(k x_2) = i  \varphi'(k \lambda_1 \lambda_2^{-1}  x_2), 
\quad
U_2(k x_2) =   \varphi(k \lambda_1 \lambda_2^{-1}  x_2).
\end{equation}
With this choice, and by \eqref{1stOrder}$_{4,1}$, 
the traction components $t_1$ and $t_2$ are expressed in terms of 
$\varphi$ as:
\begin{equation} \label{tphi}
t_1 = i b_3 \lambda_2^2(\lambda_1 \lambda_2^{-1} \varphi'' 
       		+  \varphi), \quad
t_2 = - b_3 \lambda_1 \lambda_2 \varphi'''  
       		+  \lambda_1^{-1} \lambda_2(\lambda_1 \lambda_2^{-1} C 
        	- b_3 \lambda_1 \lambda_2 - \rho v^2) \varphi',
\end{equation}
and Eq.\eqref{1stOrder}$_2$ reads
\begin{equation} \label{eqWithPhi1st}
b_3 \lambda_1^2 \varphi'''' 
 - (\lambda_1 \lambda_2^{-1} C
     - 2b_3 \lambda_1 \lambda_2 - \rho v^2)\varphi''
	+ (b_3 \lambda_1^2 - \rho v^2)\varphi = 0.
\end{equation}

Now a law of exponential decay is chosen for $\varphi$,
\begin{equation} \label{Phi}
\varphi(z) = A e^{-s_1z} + B e^{-s_2z}, \quad \Re(s_i)>0,
\end{equation}
for some constants  $A$ and $B$ (it is implicit in the form of 
this solution that $s_1$ and $s_2$ are distinct.)
By substitution into \eqref{eqWithPhi1st}, we see that the $s_i$ are 
roots of the following biquadratic,
\begin{align} \label{biquadratic}
& (b_3 \lambda_1^2)  s^4 
 - (\lambda_1 \lambda_2^{-1}C -2b_3 \lambda_1 \lambda_2 -\rho v^2)s^2 
  + (b_3 \lambda_1^2 - \rho v^2) = 0, \nonumber \\ 
& s_1^2 + s_2^2 
  = (\lambda_1 \lambda_2^{-1}C -2b_3 \lambda_1 \lambda_2 -\rho v^2)/(b_3 \lambda_1^2), \quad
s_1^2 s_2^2 = (b_3 \lambda_1^2 - \rho v^2)/(b_3 \lambda_1^2). 
\end{align}
The roots $s_1^2$ and $s_2^2$ of this real quadratic may be both real 
(and then they are positive because $\Re(s_i)>0$) 
or both complex (and then they are conjugate because 
\eqref{biquadratic}$_1$ is a real polynomial);
in both cases, $s_1^2 s_2^2 \ge 0$, and so by \eqref{biquadratic}$_3$,
\begin{equation} \label{intervalForV}
0 \le v \le \sqrt{b_3 \lambda_1^2/\rho}.
\end{equation} 
The upper limit of this interval corresponds to the speed of a bulk 
shear wave propagating along the $x_1$ direction.

Now the boundary conditions \eqref{BC1}, used in conjunction with 
\eqref{tphi}, and \eqref{biquadratic}$_2$, are:
\begin{equation}
(\lambda_1 \lambda_2^{-1}s_1^2+1)A
  + (\lambda_1 \lambda_2^{-1}s_2^2+1)B=0,
\quad
s_1(\lambda_1 \lambda_2^{-1}s_2^2+1)A
  + s_2(\lambda_1 \lambda_2^{-1}s_1^2+1)B=0,
\end{equation}
and the vanishing of the determinant for this linear homogeneous 
system of two equations gives the  \textit{exact secular equation}:
\begin{equation} \label{exactSecul}
[b_3(\lambda_1^2 - \lambda_2^2) - \rho v^2]\sqrt{b_3 \lambda_1^2}
+ (\lambda_1 \lambda_2^{-1} C - \rho v^2)
      \sqrt{b_3 \lambda_1^2 - \rho v^2} =0. 
\end{equation}
In the process, we used \eqref{biquadratic}$_{2,3}$ and dropped the 
factor $s_1-s_2$.
Note that by bringing the second term of \eqref{exactSecul} to the 
right hand side and squaring, we obtain the cubic secular equation 
\cite{Dest02b}, which has spurious roots.
Now for certain stretch ratios $\lambda_1$, $\lambda_2$,
$\lambda_3$, the speed $v$ tends to zero in \eqref{exactSecul} and 
the exact \textit{bifurcation criterion} is deduced as
\begin{equation}  \label{exactCriterion}
b_3(\lambda_1^2 - \lambda_2^2) + \lambda_1 \lambda_2^{-1} C =0. 
\end{equation}
This equation defines a surface in the space of the stretch ratios 
which separates a region where the homogeneous deformations of the 
Bell half-space are always stable from a region where they might be 
unstable.
Of course, the critical stretch ratios must also satisfy the Bell 
constraint \eqref{BellLambdas}.

We now recast the secular equation for surface waves in a polynomial
form for the positive quantity $\eta$, defined by \cite{DoOg90},
\begin{equation} 
\eta = \sqrt{1 - (\rho v^2)/(b_3 \lambda_1^2)},
\end{equation}
as
\begin{equation}	\label{BellSeculEta}
f(\eta) \equiv 
	\eta^3 + \eta^2 + (\frac{C}{b_3\lambda_1\lambda_2} - 1)\eta
            - \lambda_1^{-2}\lambda_2^2 = 0.
\end{equation}
Clearly, at $\eta = 0$ (corresponding to a transverse bulk wave),
we have $f(0) = -  \lambda_1^{-2}\lambda_2^2 < 0$; 
 at $\eta = 1$ (corresponding to $v =0$), the secular equation tends
to the bifurcation criterion
$f(1) = [b_3(\lambda_1^2 - \lambda_2^2) + \lambda_1 \lambda_2^{-1}C]
          /(b_3\lambda_1^2) = 0$.

Up to this point, the setting was that of incremental surface motions
and deformations for a general Bell-constrained half-space, maintained
in a static state of arbitrary pure homogeneous triaxial stretch.
More results may actually be obtained in this general setting 
regarding the conditions of existence and the uniqueness of a Rayleigh 
wave; 
this is done elsewhere \cite{Dest03}.
We now turn our attention to two specific types of Bell materials
and compare the results obtained in plane and equibiaxial prestrains
with those obtained for rubber-like incompressible materials.

\section{Specific forms of strain energy densities}

\subsection{Simple hyperelastic Bell materials}

For \textit{simple hyperelastic Bell materials} \cite{BeHa92a}, 
the strain energy function $\Sigma_{\text{SHB}}$ is
given by
\begin{equation}
\Sigma_{\text{SHB}} = \mathcal{C}_1 (3-i_2) +  \mathcal{C}_2(1-i_3),
\end{equation}
where $\mathcal{C}_1$ and $\mathcal{C}_2$ are positive constants.
The material response functions $\omega_0$ and $ \omega_2$ and the
quantities $b_3$ and $C$ provided by \eqref{omega} and 
\eqref{coefficients} are now
\begin{equation}
\omega_0 = - \mathcal{C}_2,
\quad
\omega_2 =  \mathcal{C}_1/i_3,
\quad
b_3 = \frac{\mathcal{C}_2 +  \mathcal{C}_1\lambda_3^{-1}}
    {\lambda_2(\lambda_1 + \lambda_2)},
\quad
C = 2(\mathcal{C}_2 +  \mathcal{C}_1 \lambda_3^{-1}).
\end{equation}
In that context, the bifurcation criterion \eqref{exactCriterion} 
simplifies considerably to 
\begin{equation} \label{criterionSimple}
3 \lambda_1 - \lambda_2 = 0,
\end{equation}
which is a particularly simple linear relationship between the stretch
ratios $\lambda_1$ and $\lambda_2$.
This bifurcation criterion is universal to the whole class of simple 
hyperelastic Bell materials because it does not depend on 
$\mathcal{C}_1$, $\mathcal{C}_2$. 
This equation delimits a plane in the stretch ratios space 
($\lambda_1, \lambda_2, \lambda_3$), which cuts the
constraint plane \eqref{BellLambdas} along the straight segment
going from the point (0,0,3) to the point 
($\textstyle{\frac{3}{4}}$, $\textstyle{\frac{9}{4}}$, 0).
Moreover, the analysis below shows that the region which is stable 
with respect to incremental perturbations  
(where there exists a root $\rho v^2 >0$ to the secular equation) is: 
$3 \lambda_1 - \lambda_2 > 0$.
In Figure 1(a), the plane \eqref{criterionSimple} cuts the triangle
of the possible values for the stretch ratios \eqref{BellLambdas}
into two parts, of which the visible one is the region of 
linear surface stability of any simple hyperelastic Bell material.

\begin{figure}
\centering
\mbox{\subfigure[Region of stability]{\epsfig{figure=curveSimple.eps,
width=.37\textwidth}}
 \quad \quad \quad \quad \quad
      \subfigure[Surface wave speed]{\epsfig{figure=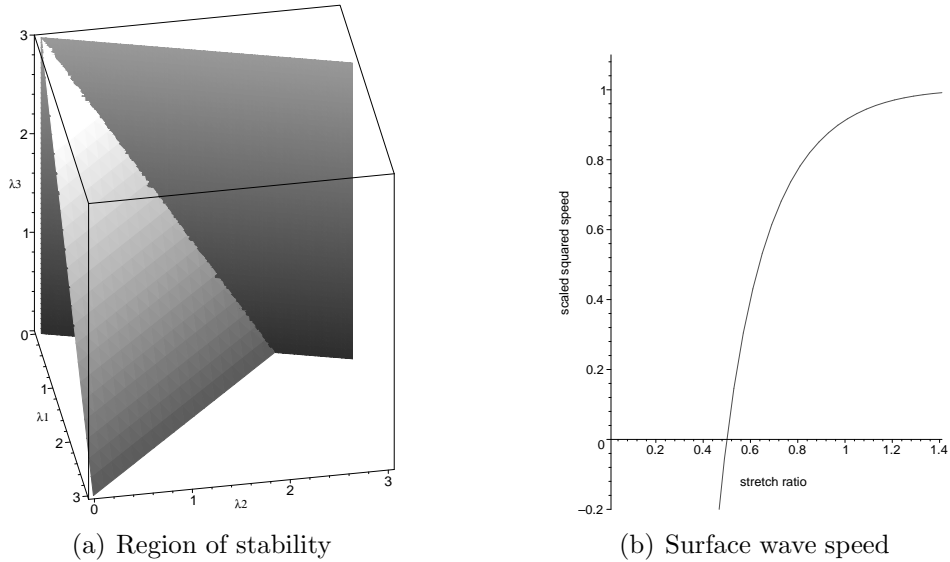,
width=.37\textwidth}}}
\caption{Near-the-surface stability for `simple hyperelastic Bell' 
materials.}
\end{figure}

For the propagating surface wave, we write the secular equation 
\eqref{BellSeculEta} in terms of $\eta$ as:
\begin{equation}	\label{simpleBellSecul}
f(\eta) =\eta^3 + \eta^2 + (1+2\lambda_1^{-1}\lambda_2)\eta 
            - \lambda_1^{-2}\lambda_2^2 = 0.
\end{equation}
As noted in the general case, $f(0) < 0$.
At the other end of the interval \eqref{intervalForV}, 
$f(1) = (3\lambda_1 - \lambda_2)(\lambda_1 + \lambda_2)\lambda_1^{-2}$.
So, because $f$ is a monotone increasing function for $\eta >0$, 
there exists a root to the secular equation \eqref{simpleBellSecul}
in the interval $[0,1]$ if and only if: 
$3\lambda_1 - \lambda_2>0$; moreover, the root is unique.

In Figure 1(b), the influence of the prestrain upon the speed of the 
surface wave is illustrated in the case of plane strain 
($\lambda_3=1$).
On the abscissa, $\lambda_1$ is increased from a compressive
value ($\lambda_1<1$) to a tensile value ($\lambda_1>1$).
The coordinate on the ordinate is the squared surface wave speed,
scaled with respect to the transverse bulk wave speed, that is 
$\rho v^2/(\mu \lambda_1^2)$.
At $\lambda_1=1$, the half-space is isotropic 
($\lambda_1= \lambda_2= \lambda_3= 1$) because of \eqref{BellLambdas}
and the scaled squared speed is equal to 0.9126, 
the value found by Lord Rayleigh \cite{Rayl85} in the incompressible 
linear isotropic case.
Under an increasing compressive load ($P_1>0$, $\lambda_1<1$), the 
surface wave speed decreases until the critical stretch of 
$(\lambda_1$)$_{\text{cr}} = 0.5$ 
(see \S\ref{Comparisons-with-incompressible-rubber}).
Conversely, under a tensile load ($P_1<0$, $\lambda_1>1$), the 
surface wave speed increases,
with the speed of the transverse bulk wave as an upper bound.

\subsection{Bell's empirical model}

For  \textit{Bell's empirical model materials} \cite{Bell85}, 
the strain energy function $\Sigma_{\text{BEM}}$ is given by
\begin{equation} \label{BellEmpirical}
\Sigma_{\text{BEM}}
 = \textstyle{\frac{2}{3}}
     \beta_0 [2(3-i_2)]^{\textstyle{\frac{3}{4}}},
\end{equation}
where $\beta_0$ is a positive constant.
The material response functions $\omega_0$ and $ \omega_2$ and the
quantities $b_3$ and $C$ provided by \eqref{omega} and 
\eqref{coefficients} are now
\begin{equation}
\omega_0 = 0,
\quad
\omega_2 =i_3^{-1} \beta_0 [2(3-i_2)]^{\textstyle{-\frac{1}{4}}}, 
\quad
b_3 = \frac{\lambda_1 \omega_2}{\lambda_1 + \lambda_2},
\quad
C = [2 - \frac{(\lambda_1 - \lambda_2)^2}{4(3-i_2)}]
    \lambda_1 \lambda_2 \omega_2.
\end{equation}
In that context, the bifurcation criterion \eqref{exactCriterion} may 
be arranged as
\begin{equation} \label{criterionEmpirical}
3 - \frac{(\lambda_1 - \lambda_2)^2}{4(3-i_2)}
-\lambda_1^{-1} \lambda_2 = 0,
\end{equation}
which is the stability equation (7.10) of Beatty and Pan 
\cite{BePa98} for this problem;
it is actually a cubic for $\lambda_1$ and $\lambda_2$.
It delimits a curved surface in the stretch ratios space 
($\lambda_1, \lambda_2, \lambda_3$), which cuts the
constraint plane \eqref{BellLambdas} into a part which is ``unstable'' 
(in the linearized theory) and a part which is always stable (in the 
linearized theory).
This partition of the triangle of possible stretch ratios is visible
on Figure 2(a).
By extension from the plane strain case where 
$\lambda_3=1$ (treated below), we deduce that the visible part of 
the triangle is the stable one.
For a clearer picture, Figure 2(b) (where the plane of the Figure 
coincides with the plane of the triangle) shows the intersection 
between the triangle and the bifurcation curve.
\begin{figure}
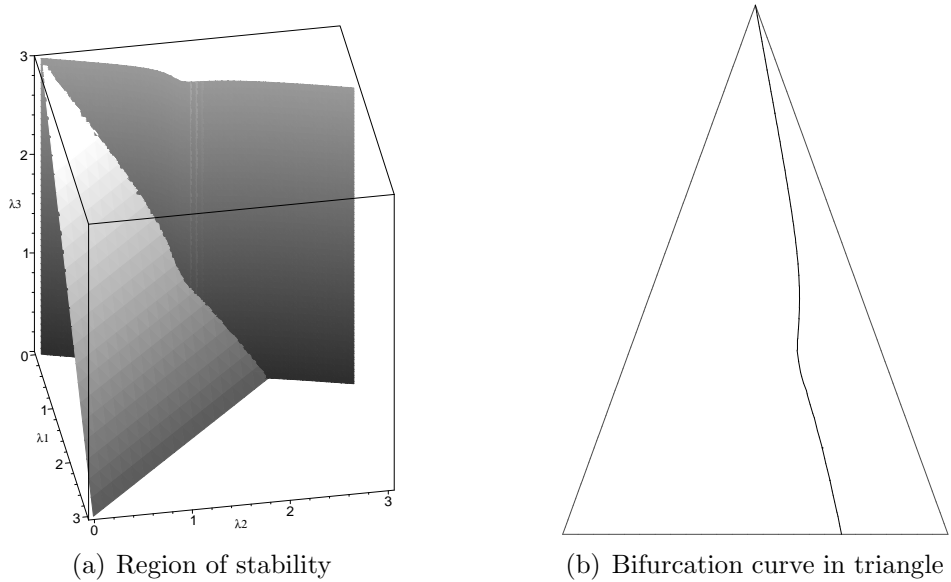

\centering
\mbox{\subfigure[Region of stability]{\epsfig{figure=curveEmpirical.eps,
width=.37\textwidth}}
 \quad \quad \quad \quad \quad
      \subfigure[Bifurcation curve in triangle]{\epsfig{figure=projectionCurve.eps,
width=.37\textwidth}}}
\caption{Near-the-surface stability for `Bell empirical model'
materials.}
\end{figure}

For the propagating surface wave, we consider only the case
where the underlying deformation is a plane strain such that 
$\lambda_3=1$. 
Then the Bell constraint \eqref{BellLambdas} reduces to 
$\lambda_1 + \lambda_2 =2$, and neither $\lambda_1$ nor $\lambda_2$ 
may be greater than 2.
Then the secular equation \eqref{BellSeculEta} reduces to:
\begin{equation}	\label{EmpiricalBellSecul}
f(\eta) =\eta^3 + \eta^2 + \lambda_1^{-1}(2-\lambda_1)\eta 
            - \lambda_1^{-2}(2-\lambda_1)^2 = 0.
\end{equation}
As noted in the general case, $f(0) < 0$.
On the other hand,  $f(1) = 2(3\lambda_1 - 2)/\lambda_1^2$,
and so, because $f$ is a monotone increasing function for $\eta>0$, 
there exists a root to the secular equation \eqref{EmpiricalBellSecul}
in the interval $[0,1]$ if and only if: 
$3\lambda_1 - 2 > 0$; moreover, the root is unique.

\subsection{Comparisons with incompressible rubber}
\label{Comparisons-with-incompressible-rubber}

The stability of a deformed half-space made of incompressible rubber
was first studied by Biot.
He used the neo-Hookean model but noted \cite[p.165]{Biot65} that the 
results were also valid for the Mooney-Rivlin model 
(Flavin \cite{Flav63} solved explicitly this latter case.)
Biot obtained the bifurcation criterion, showed that it was universal 
relative to both classes of materials, and computed the value of the 
critical stretch ($\lambda_1$)$_{\text{cr}}$ at which the rubber
half-space becomes ``unstable'' under compressive loads,
first in the case of plane strain $\lambda_3=1$, 
then in the case of the biaxial strain $\lambda_2=\lambda_3$.
Also, Green and Zerna \cite[p.137]{GrZe92} found the critical stretch 
for the biaxial prestrain $\lambda_1=\lambda_3$.
In each case a cubic must be solved in order to evaluate the critical
stretch. 
In the case of a general triaxial prestrain, the bifurcation criterion
is as follows:
\begin{equation} \label{bifurcMR}
\lambda_1^3 + \lambda_1^2 \lambda_2 + 3\lambda_1 \lambda_2^2 
 - \lambda_2^3 =0.
\end{equation}

We now show that for Bell's empirical models and for simple
hyperelastic Bell materials, the critical stretch can be found 
explicitly.

First we let the half-space made of Bell-constrained material
be deformed in such a way that there is no extension in the 
$x_3$-direction ($\lambda_3=1$).
According to \eqref{po}$_{2,3}$, this deformation is possible when the 
loads
$P_1=2(1-\lambda_1)[\lambda_1\omega_2 - \omega_0/(2-\lambda_1)]$ and
$P_3 =(1-\lambda_1)[\omega_2 - \omega_0/(2-\lambda_1)]$ are applied 
at infinity.
Then $\lambda_2 = 2 - \lambda_1$ by \eqref{BellLambdas}.
For \textit{Bell's empirical model} and for 
\textit{simple hyperelastic Bell materials}, the bifurcation 
criterion \eqref{criterionSimple} reduces respectively to 
\begin{equation}
3 (\lambda_1)_{\text{cr}} - 2 =0,
\quad \text{and} \quad
4 (\lambda_1)_{\text{cr}} - 2 =0.
\end{equation}

Then we let the half-space made of Bell-constrained material expand 
freely in the $x_3$-direction, so that $P_3=0$.
Then we have $\lambda_2 = \lambda_3 = (3-\lambda_1)/2$ 
by \eqref{po}$_3$ and \eqref{BellLambdas}, and the load
$P_1=(3/2)(1-\lambda_1)[\lambda_1\omega_2 - 2\omega_0/(3-\lambda_1)]$ 
must be applied at infinity to maintain the deformation.  
For \textit{Bell's empirical model} and for 
\textit{simple hyperelastic Bell materials}, the bifurcation 
criterion \eqref{criterionSimple} reduces respectively to 
\begin{equation}
11 (\lambda_1)_{\text{cr}} - 6 =0,
\quad \text{and} \quad
7 (\lambda_1)_{\text{cr}} - 3 =0.
\end{equation}

Finally we consider that the Bell material is subject to a biaxial 
prestrain such that $\lambda_1=\lambda_3$.
Then  $\lambda_2 = 3 - 2\lambda_1$ and 
$P_1 = P_3
   = 3(1-\lambda_1)[\lambda_1\omega_2 - \omega_0/(3-2\lambda_1)]$.
For \textit{Bell's empirical model} and for 
\textit{simple hyperelastic Bell materials}, the bifurcation 
criterion \eqref{criterionSimple} reduces respectively to 
\begin{equation}
17 (\lambda_1)_{\text{cr}} - 12 =0,
\quad \text{and} \quad
5 (\lambda_1)_{\text{cr}} - 3 =0.
\end{equation}

In Table 1, the numerical values for the critical stretches are 
given for the classes of Bell's empirical model (2nd 
column), of neo-Hookean and Mooney-Rivlin incompressible materials
\cite{Biot65, GrZe92} (3rd column), and of simple hyperelastic
Bell materials (4th column), in the cases of plane 
strain (3rd row) and of biaxial strain (2nd and 4th rows).
It appears that rubber can be compressed more than Bell's empirical 
model but less than simple hyperelastic Bell materials, 
before it loses its near-the-surface stability (in the linearized 
theory).

\begin{center}
{\footnotesize
Table 1: Critical stretch ratios $(\lambda_1)_{\text{cr}}$ 
for surface instability}
{\normalsize
\noindent
{\small
\begin{tabular}{l c c c}
\hline
\rule[-3mm]{0mm}{8mm} 
 & Bell empirical  & rubber & simple Bell
\\
\hline
$\lambda_1=\lambda_3$ & 0.706 & 0.666 & 0.600
\\
$\lambda_3=1$         & 0.667 & 0.544 & 0.500
\\
$\lambda_2=\lambda_3$ & 0.545 & 0.444 & 0.429
\\ \hline
\end{tabular}
}}
\end{center}

\subsection{A note on relative-universal bifurcation criteria}
\label{relative-universal-bifurcation-criteria}

The strain energy functions for the Mooney-Rivlin model and for the
simple hyperelastic Bell material depend both upon two distinct 
material constants:
\begin{equation}
\Sigma_\text{MR} = \mathcal{D}_1(I_1-3) + \mathcal{D}_2(I_2-3), \quad
\Sigma_\text{SHB} = \mathcal{C}_1(3-i_2) + \mathcal{C}_2(1-i_3),
\end{equation}
respectively, where $\mathcal{D}_1$, $\mathcal{D}_2$ are constants 
and $I_1$, $I_2$ are the first two invariants of the left 
Cauchy-Green tensor $\mathbf{B} = \mathbf{V}^2$.
The fact that their bifurcation criteria are `relative-universal'
\cite{Sacc01} to each class might come as a surprising result, 
but is easily understood once the strain energy functions are written
in terms of the principal stretches of the deformation \cite{DoOg90} as
$\Sigma_\text{MR}(I_1,I_2) \equiv
   W_\text{MR}(\lambda_1, \lambda_2, \lambda_3)$ and
$\Sigma_\text{SHB}(i_2,i_3) \equiv
   W_\text{SHB}(\lambda_1, \lambda_2, \lambda_3)$, where
\begin{align} \label{MR-HSB}
& W_\text{MR} = \mathcal{D}_1
                  (\lambda_1^2 + \lambda_2^2 + \lambda_3^2 - 3) 
  + \mathcal{D}_2(\lambda_1^2\lambda_2^2 + \lambda_2^2\lambda_3^2 
       + \lambda_3^2\lambda_1^2 - 3),  \nonumber \\
&W_\text{SHB} = 
   \mathcal{C}_1(3-\lambda_1\lambda_2 + \lambda_2\lambda_3 + 
                  \lambda_3\lambda_1) +
   \mathcal{C}_2(1-\lambda_1\lambda_2\lambda_3).
\end{align}

Indeed, the bifurcation criterion for a general incompressible 
hyperelastic half-space may be written in terms of the first and 
second derivatives of its strain energy function $W$ with respect 
to  $\lambda_i$ ($i=1,2$) as \cite{DoOg90},
\begin{equation} \label{generalCriterionIncomp}
\lambda_2[W_1+(2-\lambda_1^{-1}\lambda_2)W_2]
 + \lambda_1^2W_{11} - 2\lambda_1\lambda_2W_{12}
  + \lambda_2^2 W_{22} = 0.
\end{equation}
When $W$ is specialized to the Mooney-Rivlin form 
\eqref{MR-HSB}$_1$, it yields the relative-universal bifurcation 
criterion \eqref{bifurcMR}.
In fact, many subclasses of incompressible materials have a 
relative-universal bifurcation criterion.
For instance, any incompressible material with the following strain 
energy function, 
\begin{equation} \label{otherIncompr}
W = \mathcal{D}_1(\lambda_1^n + \lambda_2^n + \lambda_3^n - 3) 
  + \mathcal{D}_2(\lambda_1^n\lambda_2^n + \lambda_2^n\lambda_3^n 
       + \lambda_3^n\lambda_1^n - 3),
\end{equation}
(where $n=1,2,3,$\ldots) has the following relative-universal 
bifurcation criterion:
\begin{equation}  \label{univBifCrit}
(n-1)\lambda_1^{n+1} + \lambda_1^n \lambda_2 
  + (n+1)\lambda_1\lambda_2^n - \lambda_2^{n+1} =0.
\end{equation}
In particular, the bifurcation criterion \eqref{criterionSimple},
which happens to coincide with \eqref{univBifCrit} when $n=1$, 
is also valid for incompressible Varga materials ($n=1$ in 
\eqref{otherIncompr}).

Turning back to Bell-constrained materials, we note that it is a 
simple matter to write the quantities $b_3$ and $C$ in 
\eqref{coefficients} in terms of the derivatives of 
$W(\lambda_1, \lambda_2, \lambda_3)$.
We find that
\begin{equation}
b_3 = \frac{W_1-W_2}{\lambda_2 \lambda_3 (\lambda_1^2 - \lambda_2^2)},
\quad
C = (W_{11} - 2 W_{12} + W_{22})/\lambda_3,
\end{equation}
so that the bifurcation criterion \eqref{exactCriterion} for 
Bell materials is rewritten as
\begin{equation} \label{generalCriterionBell}
W_1 - W_2 + \lambda_1(W_{11} - 2 W_{12} +  W_{22}) = 0.
\end{equation}
When  $W$ is specialized to the simple hyperelastic 
Bell model \eqref{MR-HSB}$_2$,
it yields the relative-universal bifurcation criterion 
\eqref{criterionSimple}.
Similarly, when  $W$ is specialized to the strain energy function
of the Bell empirical model \eqref{BellEmpirical}, written as
\begin{equation}
W_\text{BEM} = \textstyle{\frac{2}{3}} \beta_0 
 [2(3-\lambda_1\lambda_2 + \lambda_2\lambda_3 + 
                  \lambda_3\lambda_1)]^{\textstyle{\frac{3}{4}}},
\end{equation}
it yields the bifurcation criterion \eqref{criterionEmpirical}.

\section{Concluding remarks on stability and on the Bell constraint} 

A word of caution is needed, in conclusion, regarding the notion of 
instability.
Throughout the paper, care was taken to talk of instability
`in the linearized theory' and of a region in the stretch ratios space
where the deformed half-space `might be unstable'.
This is so because conclusions about the actual stability of a 
finitely deformed half-space do not necessarily come out of the 
dynamical method of surface wave analysis.
As Chadwick and Jarvis \cite{ChJa79} pointed out, `the exponential
growth of a solution obtained (\ldots) on the basis of a linearized
theory eventually violates the assumption underlying the 
linearization'.
Some authors have linked instability analysis and bifurcation theory,
but as Guz remarked \cite[p.268]{Guz99}, such a comparative analysis 
`is only qualitative and to a certain extend sketchy'.
Finally, Biot \cite{Biot65} adopted a static approach to the problem
and found that at the critical stretch, the surface deflection 
(i.e. the component of the displacement normal to the surface)
became infinite; 
however, it is clear that such a deformation can hardly be called
`incremental'.
Nevertheless, it is comforting to remark that each approach yields
the same result for the critical stretches, 
and to know that, to some extent, concording experimental results 
exist \cite{BiOR61}.
On the other hand, the linear theory of stability has its limits. 
As kindly pointed out by a referee, Chadwick and Jarvis \cite{ChJa79} 
go on to say that in a situation where exponential growth occurs, 
`further enquiry is need to discover whether or not the terms 
initially neglected cause the motion to be stabilized.' 
Such a line of inquiry has indeed been followed since (see for 
instance Fu \cite{Fu93} or Fu and Rogerson \cite{FuRo94}), with the 
result that in certain cases, terms neglected in the linear theory do 
indeed contribute to a greater stability.

Finally, the motivation for the use of the Bell constraint is now 
exposed.
Using his countless experiments, James F. Bell produced a large 
literature corroborating the existence for certain metals of the 
constraint that now bears his name.  
A thorough background and detailed account of his research is given 
in a recent review by Beatty \cite{Beat01}.
Note however that the actual existence of `Bell materials' is 
controversial and that Bell's results have been criticized 
\cite{SeDo90, Stei03}. 
From a \textit{theoretical} point of view, there is a justification in 
studying classical problems (such as the one presented here) for a 
material subject to an internal constraint other than 
incompressibility.
Indeed incompressibility is an \textit{exceptional} isotropic 
constraint because the corresponding reaction stress is spherical. 
This property distinguishes incompressibility among generic isotropic 
constraints. 
For instance Pucci and Saccomandi \cite{PuSa96} proved that the 
deformation \cite{Carr67},
\[
x_1 = A X_1 + \sin \lambda X_2,
\quad
x_2 = D X_2, 
\quad 
x_3 = A X_3 - \cos \lambda X_2,
\]
where $A$, $D$, $\lambda$ are constants, is universal for all 
isotropically constrained materials, \textit{apart} from 
incompressible materials.
Also, it is always possible to subject a constrained material 
successively to a triaxial stretch followed by a simple shear,
\[
x_1 = \lambda_1 X_1 + k  \lambda_2 X_2,
\quad
x_2 = \lambda_2 X_2, 
\quad 
x_3 = \lambda_3 X_3,
\]
where the $\lambda_i$ and $k$ are constants, in such a way that the 
following relations for the Cauchy stress components are universal, 
$\sigma_{11} = \sigma_{22}$ and $\sigma_{12} = 0$, \textit{except} 
when the reaction stress is spherical (these and related points are 
developed in detail by Saccomandi in Refs.\cite{Sacc01} or 
\cite{Sacc01b}.) 
Therefore, a better understanding of the (theoretical) behaviour of 
Bell constrained materials gives us a better understanding of 
isotropic constraints and of their general -- and not exceptional -- 
mechanical properties.

\bigskip


\newpage

\bigskip

\noindent
{\Large\textbf{List of Figures and Tables}}

\bigskip

\noindent
\textbf{Figure 1: Near-the-surface stability for 
`simple hyperelastic Bell' materials.}

\medskip
\noindent
Figure 1(a): Region of stability.

Legend on graduated axes: ``$\lambda_1$'',  ``$\lambda_2$'',  
and ``$\lambda_3$''.

\medskip
\noindent
Figure 1(b): Surface wave speed.

Legend on graduated horizontal axis: ``stretch ratio''.

Legend on graduated vertical axis: ``scaled squared speed''.

\bigskip

\noindent
\textbf{Figure 2: Near-the-surface stability for 
`Bell empirical model' materials.}

\medskip
\noindent
Figure 2(a): Region of stability.

Legend on graduated axes: ``$\lambda_1$'',  ``$\lambda_2$'',  
and ``$\lambda_3$''.

\medskip
\noindent
Figure 2(b): Bifurcation curve in triangle.

\bigskip

\noindent
\textbf{Table 1: Critical stretch ratios $(\lambda_1)_{\text{cr}}$ 
for surface instability.}


\begin{thebibliography}{99}

\bibitem{Biot91}
 M.A. Biot,
 \textit{Acoustics, Elasticity, and Thermodynamics of Porous Media:
  Twenty-One Papers by M.A. Biot}. 
 Ed: I. Tolstoy 
 (Acoustical Society of America, New-York, 1991).

\bibitem{Biot65}
 -----,
 \textit{Mechanics of Incremental Deformations},
 John Wiley, New-York (1965).

\bibitem{BeHa92a}
 M.F. Beatty and M.A. Hayes,
 J. Elasticity
 \textbf{29}, 1--84 (1992).

\bibitem{Dest02b}
 M. Destrade,
 Int. J. Nonlinear Mech.
 \textbf{38}, 809--814 (2002).

\bibitem{Rome01}
 M. Romeo,
 J. Acoust. Soc. Am.
 \textbf{110}, 59--67 (2001).

\bibitem{Bell85}
 J.F. Bell,
 Int. J. Plasticity
 \textbf{1}, 3--27 (1985).

\bibitem{BeHa92b}
 M.F. Beatty and M.A. Hayes,
 Q. Jl. Mech. Appl. Math.
 \textbf{45}, 663--709 (1992).

\bibitem{BeHa95b}
 ----- and -----,
 Z.A.M.P.
 \textbf{46}, 356--371 (1995).

\bibitem{DoOg90}
 M.A. Dowaikh and R.W. Ogden,
 IMA J. Appl. Math.
 \textbf{44}, 261--284 (1990).

\bibitem{Dest03}
 M. Destrade,
 Maths. Mech. Solids (submitted).

\bibitem{Rayl85}
 Lord Rayleigh,
 Proc. R. Soc. London 
 \textbf{A17}, 4--11 (1885).

\bibitem{BePa98}
 M.F. Beatty and F.X. Pan, 
 Int. J. Non-Linear Mech.  
 \textbf{33}, 867--906 (1998).

\bibitem{Flav63}
 J.N. Flavin,
 Q. Jl. Mech. Appl. Math.
 \textbf{16}, 441--449 (1963).

\bibitem{GrZe92}
 A.E. Green and W. Zerna,
 \textit{Theoretical Elasticity},
 Dover, New-York (1992).

\bibitem{Sacc01}
 G. Saccomandi,
 In: \textit{Nonlinear Elasticity: Theory and Applications}.
 Eds: Y.B. Fu and R.W. Ogden,
 97--134 (Cambridge University Press, London, 2001).

\bibitem{ChJa79}
 P. Chadwick and D.A. Jarvis,
 Proc. R. Soc. London 
 \textbf{A366}, 517--536 (1979).

\bibitem{Guz99}
 A.N. Guz,
 \textit{Fundamentals of the Three-Dimensional Theory of Stability
 of Deformable Bodies},
 Springer-Verlag, Berlin (1999).

\bibitem{BiOR61}
 M.A. Biot, H. Od\'e, and W.L. Roever,
 Geol. Soc. America Bull. 
 \textbf{72}, 1621--1632 (1961).

\bibitem{Fu93} 
 Y.B. Fu, 
 Proc. Roy. Soc. London 
 \textbf{A443}, 59--82 (1993).

\bibitem{FuRo94}
 Y.B. Fu and G.A. Rogerson, 
 ----- 
 \textbf{A446}, 233--254 (1994). 

\bibitem{Beat01}
 M.F. Beatty,
 In: \textit{Nonlinear Elasticity: Theory and Applications}.
 Eds: Y.B. Fu and R.W. Ogden,
 58--96 (Cambridge University Press, London, 2001).

\bibitem{SeDo90}
 H.S. Sellers and A.S. Douglas, 
 Int. J. Plast. 
 \textbf{6}, 329--351 (1990). 

\bibitem{Stei03}
 D.J. Steigmann,
 Meccanica (to appear).

\bibitem{PuSa96}
 E. Pucci and G. Saccomandi,
 Math. Mech. Solids 
 \textbf{1}, 207--217 (1996). 
 
\bibitem{Carr67}
 M.M. Carroll,  
 Acta Mech. 
 \textbf{3}, 167--181 (1967).

\bibitem{Sacc01b}
 G. Saccomandi,
 In: \textit{Topics in Finite Elasticity, 
 CISM Courses and Lectures No. 424}.
 Eds: M. Hayes and  G. Saccomandi,
 95--130 (Springer-Verlag, Vienna, 2001).



\end{thebibliography}
\end{document}